\documentclass[aps,twocolumn,prl,showpacs,superscriptaddress]{revtex4}
\usepackage{graphics}
\usepackage{epsfig}

\begin{document}

\title{ {\bf Self Organized Scale-Free Networks from Merging and Regeneration}}
\author{Beom Jun Kim}
\affiliation{Department of Molecular Science and Technology, Ajou
University, Suwon 442-749, Korea}
\author{Ala Trusina}
\affiliation{Department of Physics, Ume{\aa} University, 90187 Ume{\aa}, Sweden}
\affiliation{NORDITA, Blegdamsvej 17, 2100 Copenhagen {\O}, Denmark}
\email{trusina@tp.umu.se}
\author{Petter Minnhagen}
\affiliation{NORDITA, Blegdamsvej 17, 2100 Copenhagen {\O},
Denmark} \affiliation{Department of Physics, Ume{\aa} University,
90187 Ume{\aa}, Sweden}
\author{ Kim Sneppen}
\affiliation{NORDITA, Blegdamsvej 17, 2100 Copenhagen {\O}, Denmark}

\date{\today}

\begin{abstract}
We consider the self organizing process of merging and regeneration
of vertices in complex networks and demonstrate that
a scale-free degree distribution 
emerges in a steady state of such a dynamics.
The merging of neighbor vertices in a network may be viewed as
an optimization of efficiency by minimizing redundancy.
It is also a mechanism to shorten the distance and
thus decrease signaling times between vertices in a complex network.
Thus the merging process will in particular be relevant for 
networks where these issues related to global signaling 
are of concern.
\end{abstract}


\maketitle

The ubiquitous broad degree distribution of the real world
networks has been a matter of discussions for quite some time
(see Ref.[1]-[9]). The question as to why broad degree
distributions are observed in so many different networks, has
triggered various proposals for their dynamical evolution.
Roughly these proposals can be classified into
two main scenarios: One is related to Simon's model of human
behavior, Ref.~\cite{simon}, and was introduced in a network version
under the name ``preferential attachment" (see
Ref.~\cite{barabasi}). A related scenario is found in the
protein duplication model Ref.~\cite{vazquez-sole} which is
able to generate broad degree distributions because
duplication, to some extent, mimics preferential attachment
to neighboring nodes. Another class of models is
where a scale-free distribution appears as a
result of a balance between a modeled tendency to form hubs against an
entropic pressure towards a random network with an exponential
degree distribution. This approach includes direct attempts to
construct Hamiltonians (see Ref.~\cite{berg}, \cite{manna}),
local optimization approaches \cite{martin} as well as
generation of scale-free networks by balancing a threshold for
assigning links weighted according to exponentially
distributed binding strengths \cite{capocci}.

In this paper we are presenting a new way of obtaining the scale-free 
degree distributions $P(k) \sim k^{-\gamma}$.
The proposed mechanism is associated to the phenomena 
of aggregation with injection suggested
in the context of astrophysical systems \cite{field}.
The model describes an evolving network,
in which the main components, represented by nodes,
are capable of pairwise merging, while at the same time the
size of the network is maintained by generation of new nodes.

In real world networks one may
think of the corresponding redistribution of links as a synergetic
process associated with an increased efficiency in the linking
pattern. For example, consider the network of interconnected computers.
Since the computational power of the computers improves tremendously fast,
periodically it could become more favorable to replace two
out-dated neighboring server machines with one new machine that
can handle more connections. This
simplifies the local network topology since
the connections between the two old servers and the redundant
links to other nodes are no longer needed.
At the same time new servers may be
constantly created to fulfill new demands. The generic merging or
take-over process is defined by the update rule:
\begin{figure} \centerline{\epsfig{file=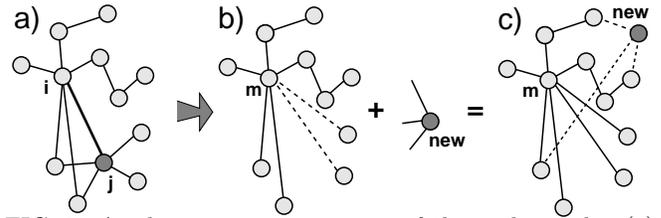,
height=0.12\textheight,angle=0}} \vskip -0.5cm
\caption{A
schematic representation of the update rule. (a) A node $i$ is
chosen at random and one of its neighbors $j$ is randomly picked
($k_i$ and $k_j$ are the degrees of $i$ and $j$, respectively).
(b) Node $m$ is a result of the pairwise merging with degree
$k_m = (k_i-1) + (k_j-1) + N_{\rm common}$, where $N_{\rm common}$ is the
number of common neighbors of $i$ and $j$ and the subtracted 1
is due to the lost common link. Node $new$ is added with
degree $k_{new}$ from a uniform distribution and it attaches links to
 $k_{new}$ random nodes (c).}
\label{fig:algorithm}
\end{figure}

\begin{itemize}
\item
At each step
we choose the node $i$ with degree $k_i$ randomly,
and then chose one of its random neighbors $j$. (See Fig.\ref{fig:algorithm}a).
\item
The nodes $i$ and $j$ are merged together and thus a
node $m$ of degree $k_{m}=(k_i - 1) + (k_j -1) - N_{common}$ appears instead,
with $N_{common}$ being the number of nodes that are neighbors to
both $i$ and $j$.
\item
At the same time a new node of some degree $k_{new}$ is added
to the network (Fig.~\ref{fig:algorithm}b) with the links attached to $k_{new}$
random nodes (see Fig.\ref{fig:algorithm}c).
The degree $k_{new}$ of a newly added node is a random number
$r$ picked from a uniform distribution with average $\langle r \rangle$.
\end{itemize}

Effectively this update reads:
\begin{equation}
\left \{ \begin{array}{ll}
        k_i   \rightarrow  k_i  + k_j - N_{\rm  common} - 2\\
        k_{j} \rightarrow  r
  \end{array},
\right.
\label{eq:algorithm}
\end{equation}
where in addition the $N_{common}$ common neighbors are loosing one connection each,
and $r$ random nodes get one connection each.
After the merging, $i$ and $j$ lose their
identities and thus Eq.~(\ref{eq:algorithm}) can equally be
written with $i$ and $j$ exchanged.

\begin{figure}
\centerline{\epsfig{file=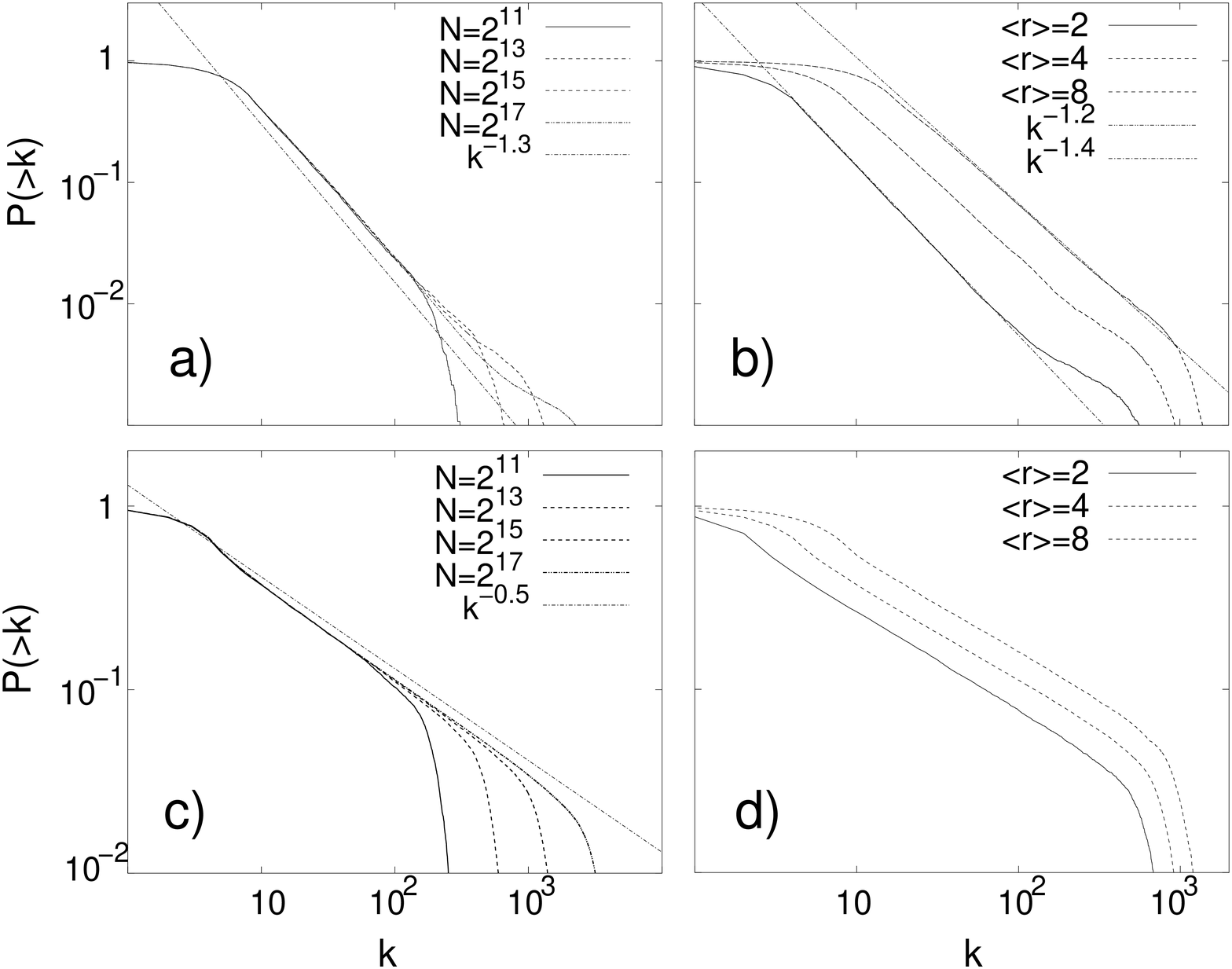, height=0.27\textheight,angle=0}}
\caption{(a) The cumulative node degree distribution $P(>k)$ for
 networks of sizes $N = 2^{11}$, $2^{13}$, $2^{15}$ and  $2^{17}$ with
the average $\langle r \rangle=8    $. The fit is the power-law
form $P(>k)\sim k^{-\gamma + 1}$ with $\gamma=2.3$. (b)
$P(>k)$ for $\langle r \rangle=2,4$
and 8 and system size $N=2^{14}$ . The straight lines are the power-law
fit with  $\gamma=2.4$ for $\langle r \rangle=2$ and
$\gamma=2.2$ for $\langle r \rangle=8$.
(c) The cumulative degree distribution  $P(>k)$ for
four different system sizes for the realization when two randomly
selected nodes are merging.
The fit is $P(>k)\sim k^{-\gamma + 1}$ with $\gamma=1.5$.
(d) $P(>k)$ for the merging of randomly selected nodes for $\langle r \rangle=2,4$,
and 8 and system size $N=2^{14}$. The slope for every $\langle r \rangle $ is  $\gamma=1.5$.
}
\label{fig:L-k}
\end{figure}

In Fig.~\ref{fig:L-k}(a)  we show the cumulative degree
distributions $P(>k)$ resulting from the update rule (1), which is the probability of
finding a vertex with degree larger than $k$, for networks of
different sizes at a steady state.  The distribution is broad, and
in fact clearly exhibits a broad range of power-law behavior
from  degree of about
$k= \langle r \rangle$ up to a cutoff which increases with system size
as shown in Fig.~\ref{fig:L-k}(a). The
crucial point to note is that the scale-free network is an emergent
property based on a simple merging process and that the driving mechanism is not related
to preferential attachment.

In order to clarify this further we first note that the present
neighbor-merging process (see Fig.~\ref{fig:L-k}) in some sense implicitly
introduces a touch of "preferential" since, when taking a random
neighbor of a random node, the neighbor is  in some
average sense selected with a probability proportional to its
degree. However, this touch of "preferential" is not  an
essential part of why the merging generates scale-freeness as is
illustrated by considering a version of the merging process where
two random nodes are merged irrespective of whether or not they
are connected, i.e. without any touch of "preferential".
In that case one always, independently of the value of $\langle r \rangle$,
obtains a scale-free distribution $P(>k) \propto 1/k^{0.5}$ (see Fig.\ \ref{fig:L-k}c, d).
Thus it is the merging, and not the preferential attachment that is the primary cause of
the scale-free distribution.
In fact it is remarkable that the neighbor-merging produces a narrower
distribution than the completely random merging. (Compare Fig.~\ref{fig:L-k}b and
Fig.~\ref{fig:L-k}b where $\gamma \sim 2.3$ for the neighbor-merging and $\gamma=1.5$ for the
random merging.)
This reflects the property of merging to limit growth of hubs by their absorption
of singly connected neighbors.
(See the update rule (1), if $k_j=1$ then $k_i \rightarrow k_i -1$.)
This tendency is stronger in the neighbor-merging process than in the
random node merging due to the
larger probability for a hub to merge with a single node in the former case.
It means that the ''touch of preferential'' for the neighbor-merging actually inhibits the growth
of a hub. This is in fact opposite to case of ''preferential attachment'' where hubs are
thriving by accumulation of neighbors of low degree. 
\begin{figure}
 \centerline{\epsfig{file=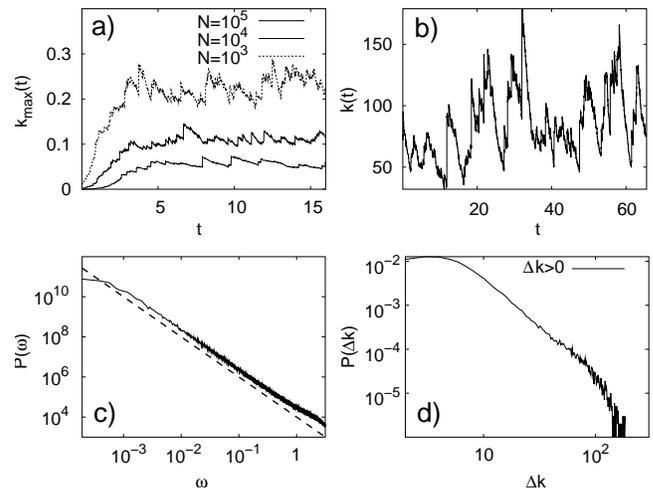, height=0.27\textheight,angle=0}}
\caption{Dynamics of the model network, with time
measured as the number of updates per node.
(a) Transient evolution of the maximal degree $k_{max}$
in the system for different system sizes.
Notice that the y-axis is normalized with system size N.
(b) Degree $k(t)$ of a given random node as a function of time $t$. When the
 node is merged, $k(t)$ shows a sudden abrupt increase.
(c) The power spectrum $P(\omega)$ versus the angular frequency
$\omega$ obtained from the Fourier transformation of $k(t)$.
The fit is a power-law with exponent $-2$.
(d) The distribution of changes $\Delta k$ for the involved
nodes in each update.
}
 \label{fig:fft}
\end{figure}

In the following we will discuss the original formulation of the mechanism ~(\ref{fig:algorithm}).
The main motivation being that the neighbor-merging version is likelier to be relevant for real
networks
since merging among the neighbors seems more natural
than the merging of random nodes.
In that case the only parameter in the system is the average degree of the nodes,
set by the average value of $\langle r \rangle$.
In Fig.~\ref{fig:L-k}(b) we show $P(>k)$ for three different
values of $\langle r \rangle$. The exponent $\gamma$ in the  power-law
form $P(>k) \sim k^{-\gamma + 1}$ decreases 
as one increases $\langle r \rangle$.  For instance, $\gamma=2.4$,
 and $2.2$ for $\langle r \rangle=2$ and $8$ respectively.
Furthermore we verified that in all cases
the steady state degree distribution depends neither on the initial average degree
nor on the shape of the initial degree distribution, be it a
narrow distribution (star-like or exponential) or a broad one (scale-free). 


An additional noteworthy feature of the model is that it
produces networks without any degree-degree correlations.
We measure the correlation profile
$C(k_1,k_2)\equiv N(k_1,k_2)/N(k_1,k_2)_{\rm randomised} - 1$,
 where $N(k_1, k_2)$ is the number
of edges connecting nodes with degrees $k_1$ and $k_2$, and
$N(k_1,k_2)_{\rm randomised}$ is the
corresponding quantity measured in the randomized
network by many steps of edge
exchanges (see Ref.~\cite{sneppen} for details). We always find
$|C(k_1, k_2)|$ to be small, $|C| < O(10^{-1})$, implying
absence of significant degree-degree correlation in the networks
emerging from the update rule~(\ref{eq:algorithm}).

The emergence of scaling is associated with a transient during which larger
hubs are slowly forming, resulting in a self sustaining ecology
with a broad degree distribution.
This transient is illustrated in Fig.~\ref{fig:fft}(a),
where we follow the degree of the, at any time, most connected node in the system.
This allows us to follow the transient approach towards the steady state.
By data collapse (not shown) we found that the transient time increases
slightly with system size, $\propto N^{0.2}$,
whereas the maximum connected node at steady state has a degree,
$k_{max} \propto N^{0.3}$.
In Fig.~\ref{fig:fft}(b) we follow a single node in steady state for a $N=10^3$,
and observe an intermittent behavior, which as seen in Fig.~\ref{fig:fft}(c),
can be characterized by a $1/\omega^2$ power spectrum.
The power-law decay form of the power
spectrum indicates the absence of a characteristic time scale,
which is in parallel with the absence of the
characteristic degree scale in the limit of large $N$  observed in
Fig.~\ref{fig:L-k}.
We stress, that although the $1/\omega^2$ spectra
resembles the one obtained for a random walk process,
the actual dynamics is richer. This is reflected
by the large jumps in increases of degree $k(t)$ in Fig.~\ref{fig:fft}(b).
This is
quantified further by the broad distribution
of changes $P(\Delta k)$ in Fig.~\ref{fig:fft}(d).
\begin{figure}
\centerline{\epsfig{file=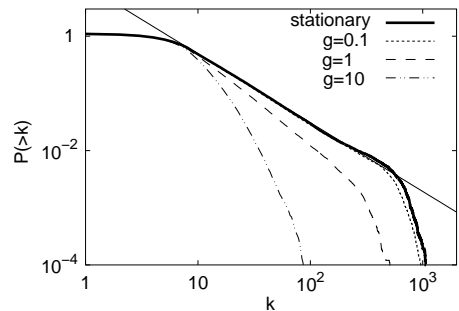, height=0.17\textheight,angle=0}}
\caption{The cumulative degree distribution for the growing network.
Different curves correspond to different rates of growth.
The solid curve is the degree distribution in non-growing case,
and the line fit has slope $\gamma=-2.2$.
With increasing the growth rate the distribution deviates from the stationary
distribution:
For moderate growth rates ($g=0.1$ and 1) the distribution remains scale-free,
whereas it collapses to exponential for larger growth rates ($g=10$).}
\label{fig:rate}
\end{figure}

So far we have been discussing a non-growing version
of the network with the number of nodes being
constant at each time step. One might argue that the
majority of the real world networks are not in a steady
state, but increase in size.
For example, both the World Wide Web and the Internet are growing.
Our merging algorithm  can be extended to include a growth process
if we add new nodes at each time at rate higher than that of merging.
We stress that the growth is {\em non-preferential} in the sense that
the newly added nodes link to the existing nodes
with a probability that is
independent of their degree. We start from a
small initial network
and grow it with $\langle r\rangle=4$ at various values
of the growth rate $g$ until
the network sizes reach $N=10^5$. In Fig.~\ref{fig:rate}
we show $P(>k)$ at the growth rates 0.1, 1.0, and 10.
For example, if the growth rate is 0.1, one vertex is added
per every 10 steps on average.  If the network size increases
very slowly, say one node per hundred basic steps,
then the degree distribution approaches the one obtained
for the non-growing case. As the growth rate is increased
the distribution still retains its power-law form shape,
but the slope $\gamma$ increases to,  e.g., 2.8
for the growth rate 1.
As the growth rate is further increased, $\gamma$ reaches 3, and then
the power-law form begins to break down, and the degree distribution
turns into  the exponential one.
The change in the slope reflects the difference in merging frequency and
the frequency at which new nodes (typically nodes of low degree
$k\sim \langle k \rangle $) are added to the system. In other words,
the competition between the two time scales, one related with the merging
and the other related with the growth, results in different degree
distributions as the growth rate is changed.
The overall feature is that the  degree distribution becomes narrower at a higher growth rate
because there is not enough time for the merging of the newly
added nodes to spread across the whole network before the system grows further.


We also note that the fact that the merging and the regeneration 
mechanism gives rise to scale-free distributions does not hinge 
on the network structure per se.
It is also applicable to entities characterized by just a scalar number, as is further
discussed in \cite{ourSymmetricCase}.

In this Letter we propose a generic and robust mechanism for
obtaining a broad, scale-free, degree distribution in
networks where merging of nodes
play a major role. The mechanism
differs fundamentally from the
preferential attachment mechanism \cite{albert}
where a broad distributions are generated during
gradual growth of hubs.
The broad distribution resulting from merging and regeneration process emerges after a
transient with slow building up of a zoo of nodes of various
degrees which, as the steady state is approached, together build
up a scale-free distribution.
We suggest that the mechanism could be relevant in a number of
real world networks where the redistribution of links is
associated with increasing efficiency in the linking pattern through
minimization of pathway lengths.
As an example we suggest that merging may be the
effective result of evolution of
architecture of protein regulatory networks
in a cell.
In these transcription and signalling networks,
the time it takes to transmit signals is important ~\cite{alon},
and it may thus be advantageous to eliminate an
intermediate regulatory protein and move its regulation
to an upstream regulatory protein. With the addition of new functions in form
of new proteins (nodes),  this effectively corresponds
to the merging and regeneration model proposed in this paper.

\vspace{0.5cm}

\acknowledgements
A.T., P.M. and K.S. acknowledges the support of Swedish Research Council
through Grants No. 621 2002 4135 and 629 2002 6258.
B.J.K acknowledges the support by Korea Science and Engineering Foundation
through Grant No. R14-2002-062-01000-0.

\end{document}